\documentclass[aps,pra,amssymb,amsmath,eqsecnum,nofootinbib]{revtex4}
\usepackage{graphicx}

\newtheorem{definition}{Definition}[section]
\newtheorem{theorem}{Theorem}[section]
\newtheorem{proposition}{Proposition}[section]
\newtheorem{lemma}{Lemma}[section]

\newtheorem{remark}{Remark}[section]

\newenvironment{hypothesis}{HP: \begin{center}} {\end{center}}
\newenvironment{thesis}{TH: \begin{center}} {\end{center}}
\newenvironment{proof}{\begin{center}PROOF: \end{center}} {$ \blacksquare $}
\newtheorem{example}{Example}[section]
\begin{document}
\title{Renormalization Group versus Kolmogorov-Sinai entropy: a very simple remark}
\author{Gavriel Segre}
\email{Gavriel.Segre@msi.vxu.se}
 \affiliation{International Center
for Mathematical Modelling in Physics and Cognitive Sciences,
University of V\"{a}xj\"{o}, S-35195, Sweden}
\begin{abstract}
 A very simple remark concerning a link between the notions of
 Kolmogorov-Sinai entropy and of Renormalization Group is
 performed.
\end{abstract}
\maketitle
\newpage
\section{Introduction}
Kadanoff-Wilson's \footnote{As it often happens the attribution of
paternity is a subtle matter. In particular a little dispute
exists as to the contribution by C. Di Castro and G. Jona-Lasinio.
As it is more appropriate in these cases one has to listen both
the viewpoints \cite{Benfatto-Gallavotti-95}, \cite{Fisher-99}.}
renormalization group \cite{Binney-Dowrick-Fisher-Mewmann-92}, and
the Kolmogorov-Sinai's entropy \cite{Sinai-94} have one
similarity: they involve a sequence of partitions of the
underlying probability space respectively decreasing and
increasing with respect to the coarse-graining ordering relation.

In the introduction of \cite{Benfatto-Gallavotti-95} one, indeed,
reads that among the applications of renormalization-group there
is the analysis of the onset of chaos in dynamical systems.

Both therein and, as far as as I know, elsewhere, anyway (also
taking into account M.J. Feigenbaum's stuff \cite{Cvitanovic-84}),
a structural analysis concerning the inter-relation between
Renormalization Group and the Kolmogorov-Sinai entropy is, at
least as far as I know, still lacking.

These brief notes are intended (to try) to make a (very  little)
step in such a direction.
\newpage
\section{Coarse graining flows, refinements' flows and their limit points}
Let $ ( X , \sigma , \mu ) $ be a classical probability space and
let us introduce the following:
\begin{definition} \label{def:partition of a classical probability space}
\end{definition}
PARTITIONS OF $ ( X , \sigma , \mu ) $:
\begin{multline}
    {\mathcal{P}} ( X , \sigma , \mu ) \; := \; \{ P = \{ A_{i} \}_{i=1}^{n(P)} \, : n(P) \in  {\mathbb{N}}_{+} \; , \;  A_{i} \in \sigma  \; i=1 , \cdots, n(P), \; \\
      A_{i}  \, \cap \, A_{j} \, = \, \emptyset \; i,j =1 , \cdots, n(P) \, : i \neq j  \; , \; \mu (X - \cup_{i=1}^{n(P)} A_{i}) = 0  \}
\end{multline}

\smallskip

\begin{remark}
\end{remark}

Beside its abstract, mathematical formalization, the definition
\ref{def:partition of a classical probability space} has a precise
operational meaning.

Given the classical probability space $  ( X , \sigma ,  \mu ) $
let us suppose to make an experiment on the probabilistic universe
it describes using an instrument whose resolutive power is limited
in that it is not able to distinguish events belonging to the same
atom of a partition $ P = \{ A_{i} \}_{i=1}^{n} \in \mathcal{P} (
X , \sigma , \mu ) $.

Consequentially the outcome of such an experiment will be a number
\begin{equation}
  r \in \{ 1 , \cdots , n \}
\end{equation}
specifying the observed atom $ A_{r} $ in our coarse-grained
observation of $ ( X , \sigma , \mu ) $.

We will call such an experiment an \emph{operational observation
of $ ( X , \sigma , \mu ) $ through the partition P} or, more
concisely, a \emph{P-experiment}.

\smallskip

The probabilistic structure of the operational observation of $ (
X , \sigma , \mu ) $ through a partition   $ P \in {\mathcal{P}} (
X , \sigma , \mu ) $ is enclosed in the following:
\begin{definition}
\end{definition}
PROBABILITY MEASURE OF THE P-EXPERIMENT:
\begin{equation*}
    \mu_{P} \; := \; \mu |_{\sigma (P)}
\end{equation*}
where $ \sigma (P) \subset  \sigma $ is the $ \sigma$-algebra
generated by P.

Given $ P_{1}, P_{2} \in {\mathcal{P}} ( X , \sigma , \mu ) $:
\begin{definition}
\end{definition}
\emph{$ P_{1} $ is a coarse-graining $ P_{2} \; ( P_{1} \leq P_{2}
) $}:

\begin{center}
 every  atom of $ P_{1} $ is the finite union of atoms of $ P_{2} $
\end{center}

\smallskip

\begin{definition}
\end{definition}
\emph{coarsest refinement of $ A = \{ A_{i} \}_{i=1}^{n} $ and  $
B = \{ B_{j} \}_{j=1}^{m} \in {\mathcal{P}}(  X , \sigma ,  \mu )
$}:
\begin{equation}
  \begin{split}
      A \, & \vee \, B \; \in {\mathcal{P}}( X  , \sigma ,  \mu )  \\
      A \, & \vee \, B \; := \; \{ \, A_{i} \, \cap \, B_{j} \, \;  i =1 , \cdots , n \; j = 1 , \cdots , m  \}
  \end{split}
\end{equation}

One has that:
\begin{theorem}
\end{theorem}
\begin{center}
  $ \leq $ is an  ordering relation over $ {\mathcal{P}}( X  , \sigma ,
  \mu )$
\end{center}

\smallskip

Let us now introduce the following:
\begin{definition}
\end{definition}
ENTROPY OF $ P = \{ A_{i} \}_{i=1}^{n(p)} \in {\mathcal{P}} ( X ,
\sigma , \mu ) $:
\begin{equation}
    H(P) \; := \; - \sum_{i=1}^{n(P)} \mu_{P}( A_{i} ) \log_{2} \mu_{P}( A_{i} )
\end{equation}

\smallskip

\begin{remark}
\end{remark}
The entropy H(P) of the partition P measures the amount of
information that one acquires realizing the P-experiment.

\smallskip

\begin{definition}
\end{definition}
$ d :  {\mathcal{P}} ( X , \sigma , \mu ) \times  {\mathcal{P}} (
X , \sigma , \mu ) \mapsto [0, + \infty) $:
\begin{equation}
    d ( P_{1} , P_{2} ) \; := \; | H(P_{1}) -  H(P_{2})|
\end{equation}

\smallskip

\begin{remark}
\end{remark}
Let us observe that d is not a metric over $ {\mathcal{P}} ( X ,
\sigma , \mu ) $ since $ d( P_{1} , P_{2} ) = 0 \, \nRightarrow
P_{1} = P_{2} $.

\smallskip

Let us introduce the following:
\begin{definition} \label{def:coarse-graining flow}
\end{definition}
\emph{coarse-graining flow over  $ ( X , \sigma , \mu ) $}

a sequence $ \{ P_{n} \}_{n \in {\mathbb{N}}} $ such that:
\begin{equation}
    P_{n} \in {\mathcal{P}} ( X , \sigma , \mu ) \; and \; P_{n+1} \leq P_{n} \; \; \forall n \in
    {\mathbb{N}}
\end{equation}

\begin{definition} \label{def:refinements' flow}
\end{definition}
 \emph{refinements' flow over $ ( X , \sigma , \mu ) $}:

a sequence $ \{ P_{n} \}_{n \in {\mathbb{N}}} $ such that:
\begin{equation}
    P_{n} \in {\mathcal{P}} ( X , \sigma , \mu ) \; and \; P_{n} \leq P_{n+1} \; \; \forall n \in
    {\mathbb{N}}
\end{equation}

Given an arbitrary sequence of mathematical objects $ \{ a_{n} \}
_{n \in {\mathbb{N}}} $ let us introduce the following:
\begin{definition}
\end{definition}
\emph{reverse of $ \{ a_{n} \} _{n \in {\mathbb{N}}} $:}
\begin{equation}
  reverse ( \{ a_{n} \} _{n \in {\mathbb{N}}}) \; = \;  \{ b_{-n} \} _{n \in
  {\mathbb{N}}} \; : \; b_{-n} := a_{n}
\end{equation}
One has clearly that:
\begin{proposition} \label{prop:reverse inverts coarse-graining and refinements' flows}
\end{proposition}
\begin{enumerate}
    \item $ \{ P_{n} \}_{n \in {\mathbb{N}}} $ is a \emph{coarse-graining
    flow} $ \Rightarrow \; reverse (  \{ P_{n} \}_{n \in
    {\mathbb{N}}}) $ is a \emph{refinements' flow}
    \item $ \{ P_{n} \}_{n \in {\mathbb{N}}} $ is a \emph{refinements'
    flow}  $ \Rightarrow \; reverse (  \{ P_{n} \}_{n \in
    {\mathbb{N}}}) $  is a \emph{coarse-graining
    flow}
\end{enumerate}

Given a \emph{refinements' flow} or a \emph{coarse-graining flow}
$ \{ P_{n} \}_{n \in {\mathbb{N}}} $ and a partition $ P_{0} \in
{\mathcal{P}} ( X , \sigma , \mu ) $:
\begin{definition} \label{def:limit point of a refinements' flow or a coarse-graining flow}
\end{definition}
$ P_{0} $ IS A LIMIT POINT OF $ \{ P_{n} \}_{n \in {\mathbb{N}}}
$:
\begin{equation}
    \forall \epsilon > 0 \, , \, \exists N \in {\mathbb{N}} \; :
    \; d( P_{n} , P_{0} ) < \epsilon \, \forall n > N
\end{equation}

Let us observe that:
\begin{proposition} \label{prop:monotone behavior of entropy of refinements and coarse-graining flows}
\end{proposition}
\begin{enumerate}
    \item $ \{ P_{n} \}_{n \in \mathbb{N}}
$ is a \emph{refinements' flow } $ \Rightarrow \;  H(P_{n}) \leq H
( P_{n+1}) \; \; \forall n \in \mathbb{N} $
    \item  $ \{ P_{n} \}_{n \in \mathbb{N}}
$ is a \emph{coarse-graining flow } $ \Rightarrow \;  H(P_{n+1})
\leq H ( P_{n+1}) \; \; \forall n \in \mathbb{N} $
\end{enumerate}

\newpage
\section{Kolmogorov-Sinai entropy versus refinements' flows}
Let us start from the following:
\begin{definition}
\end{definition}
\emph{classical dynamical system} :

a couple $ ( ( X , \sigma , \mu) , T) $ such that:
\begin{itemize}
    \item $ ( X , \sigma , \mu)  $ is a classical probability
    space
    \item $ T : X \mapsto X $ is such that:
\begin{equation}
    \mu \circ T^{- 1} \; = \; \mu
\end{equation}
\end{itemize}

Given  a classical dynamical system $  CDS \, := \, ( (X \, ,
\sigma , \mu)  ,  T ) $, the $T^{-1}$-invariance of $ \mu $
implies that the partitions $ P = \{ A_{i} \}_{i=1}^{n} \in
\mathcal{P} ( X , \sigma , \mu ) $ and $ T^{-1}P $ have equal
probabilistic structure. Consequentially the \emph{P-experiment}
and the \emph{ $T^{-1}P $-experiment} are replicas, not
necessarily independent, of the same experiment made at successive
times.

In the same way the \emph{$ \vee_{k=0}^{n-1} \, T^{-k} P
$-experiment} is the compound experiment consisting in n
replications $ P \, , \, T^{-1} P \, , \, , \cdots , \,
T^{-(n-1)}P $ of the experiment corresponding to $ P \in
{\mathcal{P}}(X , \sigma , \mu) $.

The amount of classical information for replication we obtain in
this compound experiment is clearly:
\begin{equation*}
  \frac{1}{n} \, H(\vee_{k=0}^{n-1} \, T^{-k} P )
\end{equation*}
It may be proved (cfr. e.g. the second paragraph of the third
chapter of \cite{Kornfeld-Sinai-00}) that when n grows this amount
of classical information acquired for replication converges, so
that the following quantity:
\begin{equation}
  h( P , T ) \; := \; lim_{n \rightarrow \infty} \,  \frac{1}{n} \, H(\vee_{k=0}^{n-1} \, T^{-k} P )
\end{equation}
exists.

In different words, we can say that $ h( P , T ) $ gives the
asymptotic rate of production of classical information for
replication of the P-experiment.

\begin{definition} \label{def:Kolmogorov-Sinai entropy}
\end{definition}
\begin{equation}
  h_{CDS} \; := \; sup_{P \in {\mathcal{P}}(X , \sigma , \mu)} \, h( A , T )
\end{equation}
By definition we have clearly that:
\begin{equation}
  h_{CDS} \; \geq \; 0
\end{equation}
\begin{definition} \label{def:classical chaoticity}
\end{definition}
CDS IS CHAOTIC:
\begin{equation}
  h_{CDS} \; > \; 0
\end{equation}

\smallskip

By construction we have the following:
\begin{lemma} \label{lem:the refinements' flow of a dynamical system w.r.t. a partition}
\end{lemma}

\begin{hypothesis}
\end{hypothesis}
\begin{equation*}
  P \in {\mathcal{P}}(X , \sigma , \mu)
\end{equation*}
\begin{thesis}
\end{thesis}
\begin{center}
 $   \{ \vee_{k=0}^{n-1} T^{-k} P \}_{n \in {\mathbb{N}}} $ is a \emph{refinements' flow}.
\end{center}
from which it follows that:
\begin{theorem} \label{th:about the entropy of the refinements' flow of a dynamical system w.r.t. a partition having a limit-point}
\end{theorem}
\begin{hypothesis}
\end{hypothesis}
\begin{equation*}
    P \in {\mathcal{P}}(X , \sigma ,
  \mu)
\end{equation*}
\begin{center}
  $ \{ \vee_{k=0}^{n-1} T^{-k} P \}_{n \in {\mathbb{N}}}
  $ has a limit point
\end{center}
\begin{thesis}
\end{thesis}
\begin{equation*}
  h(P,T) \; = \; 0
\end{equation*}
\begin{proof}
If $ \{ \vee_{k=0}^{n-1} T^{-k} P \}_{n \in \mathbb{N}}
  $ has a limit point,  the rate of information gaining
  for replication of the P-experiment at a certain point tends to
  zero.
\end{proof}

\newpage
\section{A brief introduction to the Renormalization Group} \label{sec:A brief introduction to the Renormalization Group}
Let us introduce briefly the Kadanoff-Wilson's Renormalization
Group in a simple setting \footnote{The Renormalization Group
applies to any model of Classical Statistical Mechanics and in
particular to the situation in which the \emph{order-parameter}
lives on a space with cardinality greater than $ \aleph_{0} $; in
this case one often speaks about "Statistical Field Theory"
\cite{Parisi-88}, \cite{Itzykson-Drouffe-89a}. Since according to
the \emph{Osterwalder-Schrader axiomatization}
\cite{Glimm-Jaffe-87} (or according to the less rigorous vulgata
of Euclidean Field Theory \cite{Zinn-Justin-93}) Quantum Field
Theory reduces to Classical Statistical Mechanics (axiomatization
affected by the irreducibility of \emph{noncommutative probability
spaces} to \emph{classical probability spaces} and the related
superiority of the \emph{Haag-Kastler axiomatization} with respect
to the Osterwalder-Schrader one) the application of the
Kadanoff-Wilson Renormalization Group to Quantum Field Theory,
resulting  in the RG equation for the un-renormalized $ \Gamma_{n}
$ (where $ \Gamma( \phi)  = \sum_{n=0}^{\infty} \frac{1}{n !} \int
d x_{1}\cdots d x_{n} \Gamma_{n}( x_{1} , \cdots, x_{n}) \phi(
x_{1}) \cdots \phi( x_{n}) $ is the Legendre transform of the
logarithm of the partition function), is nothing but a particular
case of its general Statistical Mechanics' framework.}, for
instance a system of spins $ S_{i} = \pm 1 $ living on the sites
of a D-dimensional finite cubic lattice $ ( a \{-N , -N +1 ,
\cdots , 0 , \cdots , N-1 , N \} )^{D} $  and having dimensionless
hamiltonian:
\begin{multline} \label{eq:hamiltonian}
    \mathcal{H} \; := \; \beta \, H\; := \; - K_{0} \sum_{i} S_{i}  - K_{1} \sum_{<i,j>_{1}} S_{i} S_{j} - K_{2}  \sum_{<i,j>_{2}} S_{i}
    S_{j} - \cdots - K_{n} \sum_{<i,j>_{n}} S_{i} S_{j} - \cdots - K_{\infty,1} \sum_{<i,j,k >_{1}} S_{i}
    S_{j} S_{k} \\
    - K_{\infty,2} \sum_{<i,j,k >_{2}} S_{i}
    S_{j} S_{k} - \cdots - K_{\infty,n} \sum_{<i,j,k >_{n}} S_{i}
    S_{j} S_{k} - \cdots  \\
    - K_{\infty,\infty,1} \sum_{<i,j,k,r >_{1}} S_{i}
    S_{j} S_{k} S_{r} - K_{\infty,\infty,2} \sum_{<i,j,k,r >_{2}} S_{i}
    S_{j} S_{k} S_{r} - \cdots -  K_{\infty,\infty,n} \sum_{<i,j,k,r >_{n}} S_{i}
    S_{j} S_{k} S_{r} - \cdots
\end{multline}
(where $ \beta := \frac{1}{k_{B} T} $, $k_{B}$ being Boltzmann's
constant and T being the temperature) where $ < \cdots
>_{n} $ denotes spins having distance n.

Let us assume that the interaction decreases enough quickly at
large distances so that the vector $ \mathbf{K}$ of the
coupling-constants belongs to the space $ l_{2}( \mathbb{R} ) :=
\{ \{ x_{n} \}_{n \in \mathbb{N}} \: : \: x_{n} \in \mathbb{R} \,
, \, \sum_{n=0}^{\infty} | x_{n} |^{2} < + \infty \}$.

Let us now analyze how the dimensionless hamiltonian $ \mathcal{H}
( \mathbf{K} ) $ changes under a transformation which
coarse-grains the short-distance degrees of freedoms.

At this purpose let us divide the lattice $ ( a \{-N , -N +1 ,
\cdots , 0 , \cdots , N-1 , N \} )^{D} $  in cubic blocks of
linear dimension $ l a $ (with $ l \ll N )$, the generic block B
containing consequentially $ l^{D} $ spins, and let us associate
to each block B a block variable:
\begin{equation}
    S'_{B} \; := \; f ( \{ S_{i} \}_{i \in B})
\end{equation}
with, for instance:
\begin{equation}
    f ( \{ S_{i} \}_{i \in B}) \; := \; \left\{%
\begin{array}{ll}
    sign ( \sum_{i \in B} S_{i}) , & \hbox{if $  sign ( \sum_{i \in B} S_{i} ) \neq 0$ ;} \\
    S_{0}, & \hbox{otherwise.} \\
\end{array}%
\right.
\end{equation}
Introduced the function:
\begin{equation}
    P \{ S' , S \} \; := \; \prod_{B} \delta_{Kronecker} [ S'_{B}
    ,  f ( \{ S_{i} \}_{i \in B} ) ]
\end{equation}
the partition function of the system can then be expressed as:
\begin{multline}
    Z_{N} ( \mathbf{K} ) \; := \; \sum_{i \in ( a \{-N , -N +1 , \cdots , 0 , \cdots , N-1 , N \}
    )^{D}} \exp ( - \mathcal{H} ( \mathbf{K}) , \{ S_{i}\} ) \; = \;
    Z_{\frac{N}{l}} ( \mathbf{K}' )
     \; := \sum_{B} \exp ( - \mathcal{H} ( \mathbf{K}', \{ S'_{B} \} )  ) \; := \\
       \sum_{i \in ( a \{-N , -N +1 , \cdots , 0 , \cdots , N-1 , N \}
    )^{D}} \sum_{B}  P \{ S' , S \}  \exp ( - \mathcal{H} ( \mathbf{K}) ,  \{ S_{i}\}  )
\end{multline}
The passage from $ Z_{N} ( \mathbf{K} ) $ to $  Z_{\frac{N}{l}} (
\mathbf{K}' ) $ corresponds to a map into the space $ l_{2} (
\mathbb{R}) $  of the coupling constants:
\begin{equation}
  \mathbf{K}' \; = \; R_{l} ( \mathbf{K} )
\end{equation}
called a \emph{renormalization of the coupling constants}.

The \emph{renormalizations of the coupling constants} form a
semigroup:
\begin{equation}
    R_{l_{1} l_{2} } ( \mathbf{K} ) \; = \; R_{l_{1}} ( \mathbf{K} )
    \cdot  R_{l_{2}} ( \mathbf{K} )
\end{equation}
usually called the \emph{renormalization group}.

Adhering to the usual terminology we will will also refer to a
\emph{renormalization of the coupling constants} as to a
\emph{renormalization group transformation}.

Iterating a transformation $ R_{l} $ one performs a discrete-time
dynamics in the coupling-constants' space $ l_{2}( \mathbb{R} ) $
to which the whole conceptual apparatus of  Classical Dynamical
Systems' Theory applies (such as the theory of \emph{fixed-points
} and their \emph{basins of attraction}):

given $ \mathbf{K}_{\star} \in l^{2} ( \mathbb{R} ) $:
\begin{definition}
\end{definition}
 \emph{$\mathbf{K}_{\star} $ is a fixed point of the
 renormalization-group transformation $ R_{l} $}:
 \begin{equation*}
    R_{l} ( \mathbf{K}_{\star} ) \; = \; \mathbf{K}_{\star}
\end{equation*}
Given a \emph{fixed point} $  \mathbf{K}_{\star} $ of the
\emph{renormalization-group transformation} $ R_{l} $:
\begin{definition} \footnote{ Introduced the following inner product over $ l_{2}( \mathbb{R} )
$:
\begin{equation}
    ( \{ x_{n} \}_{n \in \mathbb{N}} ,  \{ y_{n} \}_{n \in
    \mathbb{N}} ) \; := \; \sum_{n=0}^{\infty} x_{n} y_{n}
\end{equation}
one has that $ ( l_{2}( \mathbb{R} ) , ( \cdot , \cdot ) ) $ in an
Hilbert space over $ \mathbb{R} $ so that the norm  $ \| \{ x_{n}
\}_{n \in \mathbb{N}} \| := \sqrt{ (  \{ x_{n} \}_{n \in
\mathbb{N}}, \{ x_{n} \}_{n \in \mathbb{N}})} $ induces the metric
$ d( \{ x_{n} \}_{n \in \mathbb{N}} , \{ y_{n} \}_{n \in
\mathbb{N}} ) := \| \{ x_{n} \}_{n \in \mathbb{N}} - \{ y_{n}
\}_{n \in \mathbb{N}} \| $ than can be used to define the notion
of limit in the usual way \cite{Reed-Simon-80}.}
\end{definition}
\emph{basin of attraction of $ \mathbf{K}_{\star} $ with respect
to $ R_{l} $}:
\begin{equation*}
    \mathcal{B}_{l} ( \mathbf{K}_{\star} ) \; := \; \{ \mathbf{K} \in
    l_{2} ( \mathbb{R} ) \, : \; \lim_{n \rightarrow + \infty}
    R_{l}^{n} ( \mathbf{K} ) \: = \:  \mathbf{K}_{\star} \}
\end{equation*}
where $   R_{l}^{n} $ denotes the $ n^{th} $-iterate of $ R_{l}$.
\smallskip

\begin{remark}
\end{remark}
Since each \emph{renormalization group transformation} $ R_{l} $
corresponds to a reduction of the number of degrees of freedom of
a factor $ l^{D} $ one could think that a renormalization group
flow necessarily terminates with the elimination of all the
degrees of freedom.

Taking the thermodynamic limit $ N \rightarrow + \infty $ it
follows that an infinite number of iterations of a
\emph{renormalization group transformation} $ R_{l} $ is required
in order to eliminate all the degrees of freedom.

It is only under the thermodynamical limit that singularities in
the free-energy $ F := - \frac{1}{\beta} \log Z $ or its
derivatives can occur.

\smallskip

\begin{remark}
\end{remark}
According to Ehrenfest's classification  a \emph{critical point of
$ n^{th} $ order} is  a point on which the free-energy is
differentiable $ n-1 $ times, but not n times. Following the usual
terminology \cite{{Parisi-88}} we will call a \emph{transition of
second order} any critical point of Ehrenfest-ordering greater or
equal than two. The phenomenon of Universality of the
long-distance behavior in the phase transitions of second order is
owed to the fact that different physical systems correspond to
different points of a same basin of attraction $ \mathcal{B}_{l} (
\mathbf{K}_{\star} )$. Such a basin of attraction is then also
called a \emph{universality class}.

\newpage

\section{Renormalization group as a particular kind of coarse-graining flow}

Let us consider the classical probability space  $ (  ( a \{-N ,
-N +1 , \cdots , 0 , \cdots , N-1 , N \} )^{D} ,
\mathcal{B}_{Borel} , \mu  ) $ where:
\begin{equation}
    d \mu ( \{ S_{i} \}_{i \in ( a \{-N ,
-N +1 , \cdots , 0 , \cdots , N-1 , N \} )^{D}} ) \; := \; N \exp[
- \mathcal{H} ( \mathbf{K} , \{ S_{i} \} )] \prod_{i \in ( a \{-N
, -N +1 , \cdots , 0 , \cdots , N-1 , N \} )^{D}} \delta (
S_{i}^{2} -1) d S_{i}
\end{equation}
where N is a normalization constant.

 The coarse-graining underlying the \emph{renormalization
group transformation}  $ R_{l} $ may be represented by the
partition $ P \in \mathcal{P} (   ( a \{-N , -N +1 , \cdots , 0 ,
\cdots , N-1 , N \} )^{D} , \mathcal{B}_{Borel} , \mu  ) $ whose
atoms are the different blocks B by which $ ( a \{-N , -N +1 ,
\cdots , 0 , \cdots , N-1 , N \} )^{D} $ has been divided.

The iteration of $ R_{l} $ corresponds to a \emph{coarse-graining
flow} $ \{ P^{(l)}_{n} \}_{n \in \mathbb{N}} $ over the classical
probability space  $ (  ( a \{-N , -N +1 , \cdots , 0 , \cdots ,
N-1 , N \} )^{D} , \mathcal{B}_{Borel} , \mu  ) $ to which
corresponds the flow of probability measures $ \{
\mu_{P^{(l)}_{n}} \}_{n \in \mathbb{N}}$.

Let us now suppose that the initial condition $ \mathbf{K} $ of
the \emph{renormalization group flow} belongs to the the basin of
attraction $ \mathcal{B}_{l} ( \mathbf{K}_{\star} ) $ of a fixed
point $ \mathbf{K}_{\star} $.

It follows that the \emph{coarse-graining flow} $ \{ P^{(l)}_{n}
\}_{n \in \mathbb{N}} $ has a limit point (according to the
definition \ref{def:limit point of a refinements' flow or a
coarse-graining flow}).

Let us now introduce:
\begin{equation}
    \{ \tilde{ P}^{(l)}_{-n} \}_{n \in \mathbb{N}} \; := \; reverse
    ( \{ P^{(l)}_{n} \}_{n \in \mathbb{N}} )
\end{equation}

By Proposition \ref{prop:reverse inverts coarse-graining and
refinements' flows} $ \{ \tilde{ P}^{(l)}_{-n} \}_{n \in
\mathbb{N}} $ is a \emph{refinements' flow}.

Let us now  consider  a $ \mu$-preserving map $ T_{l}: ( ( a \{-N
, -N +1 , \cdots , 0 , \cdots , N-1 , N \} )^{D} \mapsto ( ( a
\{-N , -N +1 , \cdots , 0 , \cdots , N-1 , N \} )^{D} $ such that:
\begin{equation}
      \vee_{k=0}^{n-1} T^{-k} \tilde{P}^{(l)}_{0} \; = \; \tilde{P}^{(l)}_{-n} \; \; \forall n \in \mathbb{N}
\end{equation}
If our knowledge of the \emph{renormalization group flow} allowed
us to know that also $ \{ \tilde{P}^{(l)}_{-n} \}_{n \in
\mathbb{N}} $ has a limit point we could use theorem\ref{th:about
the entropy of the refinements' flow of a dynamical system w.r.t.
a partition having a limit-point} to infer that:
\begin{equation}
    h( \tilde{P}^{(l)}_{0} , T_{l}) \; = \; 0
\end{equation}
\begin{example}
\end{example}
Let us consider the simplest possibility, i.e. the one dimensional
Ising model corresponding to the assumptions that the only
coupling constants different from zero are $ K_{0} $ and $ K_{1} $
and, obviously, that $D=1$.

Let us impose periodic boundary conditions $ S_{N+1} := S_{1} $

The N-spin partition function can be written as:
 \begin{equation}
    Z_{N} \; = \; Tr \mathbf{T}^{N} \; = \lambda_{+}^{N} + \lambda_{-}^{N}
\end{equation}
where $ \mathbf{T} $ is the \emph{transfer matrix}:
\begin{equation}
    \mathbf{T} \; := \; \left(%
\begin{array}{cc}
  \exp( K_{0}+K_{1}) & \exp ( - K_{1} )  \\
   \exp ( - K_{1} ) &  \exp( K_{0}-K_{1})   \\
\end{array}%
\right)
\end{equation}
and where $ \lambda_{\pm} $ are its eigenvalues:
\begin{equation}
  \lambda_{\pm} \; = \; \exp( K_{1}) [ \cosh (  K_{0})
  \, \pm \, \sqrt{ \sinh^{2} (  K_{0} ) + \exp ( - 4 \ K_{1}
  )}]
\end{equation}
Let us consider as blocks couples of nearest neighbors spins. One
has that:
\begin{equation}
    Z_{\frac{N}{2}} ( \mathbf{K}' ) \; = \; Tr \mathbf{T}'^{\frac{N}{2}}
\end{equation}
where clearly:
\begin{equation} \label{eq:first equation for the renormalized transfer matrix}
 \mathbf{T}' \; = \; \mathbf{T}^{2}
\end{equation}
Let us impose that, up to a multiplicative constant,  $
\mathbf{T}' $ has the same form as $ \mathbf{T} $:
\begin{equation}  \label{eq:second equation for the renormalized transfer matrix}
  \mathbf{T}'  \; = \; c  \left(%
\begin{array}{cc}
  \exp( K_{0}'+K_{1}') & \exp ( - K_{1}' )  \\
   \exp ( - K_{1}' ) &  \exp( K_{0}'-K_{1}')   \\
\end{array}%
\right)
\end{equation}
It is useful \cite{Huang-87} to parametrize the
coupling-constants's space introducing the vector:
\begin{equation}
  \mathbf{V} \; := \; ( V_{0} , V_{1} )
\end{equation}
where:
\begin{equation}
    V_{i} \; := \; \exp ( - K_{i} ) \; \; i=0,1
\end{equation}
The renormalization group transformation $ \mathbf{K}' = R_{2} (
\mathbf{K} ) $ induces an analogous  map $  \mathbf{V}' =
\check{R}_{2}( \mathbf{V} ) $ where:
\begin{equation}
  \mathbf{V}' \; := \; ( V_{0}' , V_{1}' )
\end{equation}
\begin{equation}
    V_{i}' \; := \; \exp ( - K_{i}' ) \; \; i=0,1
\end{equation}
The map $ \check{R}_{2} $, obtained comparing eq.\ref{eq:first
equation for the renormalized transfer matrix} with
eq.\ref{eq:second equation for the renormalized transfer matrix}
is explicitly given by:
\begin{equation}
    V_{0}' \; = \; \frac{ ( V_{1}^{4} + V_{0}^{2})^{\frac{1}{2}}}{ ( V_{1}^{4} + \frac{1}{V_{0}^{2}} )^{\frac{1}{2}}     }
\end{equation}
\begin{equation}
    V_{1}' \; = \; \frac{ ( V_{0} + \frac{1}{V_{0}})^{\frac{1}{2}}}{ ( V_{1}^{4} + \frac{1}{V_{1}^{4}} + V_{0}^{2}+\frac{1}{V_{0}^{2} })^{\frac{1}{4}}}
\end{equation}
\begin{equation}
    c \; = \; ( V_{0} + \frac{1}{V_{0}})^{\frac{1}{2}} ( V_{1}^{4} + \frac{1}{V_{1}^{4}} + V_{0}^{2}+\frac{1}{V_{0}^{2} })^{\frac{1}{4}}
\end{equation}

Let us now construct the \emph{coarse-graining flow} $
 \{ P_{n}^{(2)} \}_{n \in \mathbb{N}} $.

One has clearly that:
\begin{multline}
    P_{0}^{(2)} \; = \; \{ \{ - a N,- a(N+1) \} , \{ - a(N+2) , - a(N+3) \} , \{ -a(N+4) , - a(N+5) \}
    , \\
    \{ - a(N+6) , -a(N+7) \}  ,
    \cdots , \{  a(N-7) ,  a(N-6) \} , \{ a(N-5) , a(N-4) \} , \{ a(N -3) , a(N -
    2) \} , \{ a(N-1) , a N \}
\end{multline}
\begin{multline}
     P_{1}^{(2)} \; = \; \{ \{ - a N,- a(N+1) , - a(N+2) , - a(N+3) \} , \{ -a(N+4) , - a(N+5) , - a(N+6) , -a(N+7) \}
     , \\
    \cdots , \{  a(N-7) ,  a(N-6) ,  a(N-5) , a(N-4) \} , \{ a(N -3) , a(N -
    2) , a(N-1) , a N \} \}
\end{multline}
\begin{multline}
     P_{2}^{(2)} \; = \; \{ \{ - a N,- a(N+1) , - a(N+2) , - a(N+3)  , -a(N+4) , - a(N+5) , - a(N+6) , -a(N+7) \}
     , \\
    \cdots , \{  a(N-7) ,  a(N-6) ,  a(N-5) , a(N-4) ,  a(N -3) , a(N -
    2) , a(N-1) , a N \} \}
\end{multline}
and so on.

Introduced the \emph{refinements' flow}:
\begin{equation}
    \{ \tilde{ P}^{(2)}_{-n} \}_{n \in \mathbb{N}} \; := \; reverse
    ( \{ P^{(2)}_{n} \}_{n \in \mathbb{N}} )
\end{equation}
let us suppose to have a $ \mu$-preserving map $ T_{2}:   a \{-N ,
-N +1 , \cdots , 0 , \cdots , N-1 , N \}  \mapsto   a \{-N , -N +1
, \cdots , 0 , \cdots , N-1 , N \}  $ such that:
\begin{equation}
      \vee_{k=0}^{n-1} T^{-k} \tilde{P}^{(l)}_{0} \; = \; \tilde{P}^{(l)}_{-n} \; \; \forall n \in \mathbb{N}
\end{equation}

Let us now analyze the structure of  the \emph{renormormalization
group flow}:

performing in inverted sense  the basin of attraction of any fixed
point $ \mathbf{V}_{\lambda} := ( \lambda , 1) , \lambda \in ( 0 ,
1)$ one sees that it is a sequence converging to $ V_{\bullet} :=(
1,0 ) $.

So it follows that the associated \emph{refinements' flow}  $ \{
\tilde{P}^{(l)}_{- n} \}_{n \in \mathbb{N}} $ has \emph{a limit
point}.

Hence, by theorem\ref{th:about the entropy of the refinements'
flow of a dynamical system w.r.t. a partition having a
limit-point}, we can infer that:
\begin{equation}
    h( \tilde{P}^{(l)}_{0} , T_{l}) \; = \; 0
\end{equation}

\newpage
\section{Acknoledgements}
I acknowledge funding related to a Marie Curie post-doc fellowship
of the EU network on "Quantum Probability and Applications in
Physics, Information Theory and Biology" contract
HPRNT-CT-2002-00279 (prolonged of two months). I would like to
thank prof. A. Khrennikov for stimulating discussions; of course
he has no responsibility of any error contained in these pages.


\newpage
\begin{thebibliography}{10}

\bibitem{Benfatto-Gallavotti-95}
G.~Benfatto~G. Gallavotti.
\newblock {\em Renormalization Group}.
\newblock Princeton University Press, Princeton, 1995.

\bibitem{Fisher-99}
M.E. Fisher.
\newblock {R}enormalization {G}roup {T}heory: its {B}asis and {F}ormulation in
  {S}tatistical {P}hysics.
\newblock In T.Y. Cao, editor, {\em {C}onceptual {F}oundations of {Q}uantum
  {F}ield {T}heory}, pages 89--135. Cambridge University Press, Cambridge,
  1999.

\bibitem{Binney-Dowrick-Fisher-Mewmann-92}
J.J. Binney N.J. Dowrick A.J. Fisher~M.E. Newman.
\newblock {\em The Theory of Critical Phenomena. An Introduction to the
  Renormalization Group}.
\newblock Oxford University Press, Oxford, 1992.

\bibitem{Sinai-94}
Ya.~G. Sinai.
\newblock {\em Topics in Ergodic Theory}.
\newblock Princeton University Press, Princeton, 1994.

\bibitem{Cvitanovic-84}
P.~Cvitanovic.
\newblock {\em Universality in Chaos}.
\newblock Institute of Publishing, Bristol, 1984.

\bibitem{Kornfeld-Sinai-00}
I.P. Kornfeld~Y.G. Sinai.
\newblock General {E}rgodic {T}heory of {G}roups of {M}easure {P}reserving
  {T}rasformations.
\newblock In Y.G. Sinai, editor, {\em Dynamical Systems, Ergodic Theory and
  Applications}. Springer Verlag, Berlin, 2000.

\bibitem{Parisi-88}
G.~Parisi.
\newblock {\em Statistical Field Theory}.
\newblock Perseus Books, Reading (Massachusetts), 1988.

\bibitem{Itzykson-Drouffe-89a}
C.~Itzykson~J.M. Drouffe.
\newblock {\em Statistical Field Theory. Vol.1: from Brownian motion to
  renormalization and lattice gauge theory}.
\newblock Cambridge University Press, Cambridge, 1989.

\bibitem{Glimm-Jaffe-87}
J.~Glimm~A. Jaffe.
\newblock {\em Quantum Physics}.
\newblock Springer-Verlag, New York, 1987.

\bibitem{Zinn-Justin-93}
J.~Zinn-Justin.
\newblock {\em Quantum Field Theory and Critical Phenomena}.
\newblock Oxford University Press, New York, 1993.

\bibitem{Reed-Simon-80}
M.~Reed~B. Simon.
\newblock {\em Methods of Modern Mathematical Physics: vol.1 - Functional
  Analysis}.
\newblock Academic Press, 1980.

\bibitem{Huang-87}
K.~Huang.
\newblock {\em Statistical Mechanics}.
\newblock John Wiley and Sons, New York, 1987.

\end{thebibliography}
\end{document}